\documentclass[a4paper,titlepage,nofootinbib,12pt,oneside,final]{revtex4}
\usepackage{amsmath} \usepackage{graphicx} \usepackage{amsfonts}
\usepackage{array} \usepackage{amsthm} \usepackage{bm}

\usepackage{latexsym}
\newcommand{\e}{{\rm e}}

\newcommand{\bw}{\begin{widetext}}
\newcommand{\ew}{\end{widetext}}
\newcommand{\be}{\begin{equation}}
\newcommand{\ee}{\end{equation}}
\newcommand{\bea}{\begin{eqnarray}}
\newcommand{\eea}{\end{eqnarray}}
\newcommand{\nn}{\nonumber}

\newcommand{\s}{\sigma}

\begin{document}
\begin{flushright}DTP-MSU/07-25,
\end{flushright}

\title{Improved generating technique for $D=5$ supergravities and
squashed Kaluza-Klein Black Holes}

\author{Dmitri V. Gal'tsov} \email{galtsov@phys.msu.ru}
\affiliation{Department of Theoretical Physics, Moscow State
University, 119899, Moscow, Russia}

\author{Nikolai G. Scherbluk} \email{shcherbluck@mail.ru}
\affiliation{Department of Theoretical Physics, Moscow State
University, 119899, Moscow, Russia}

\date{\today}

\begin{abstract}
\section*{Abstract}
Recently we suggested a  solution generating technique  for
five-dimensional supergravity with three Abelian vector fields based
on the hidden $SO(4,4)$ symmetry  of the three-dimensionally reduced
theory. This technique generalizes the $G_{2(2)}$ generating
technique developed earlier for minimal 5D supergravity
(A.~Bouchareb, G.~Cl\'ement, C-M.~Chen, D.~V.~Gal'tsov,
N.~G.~Scherbluk, and Th. Wolf, Phys. Rev. D {\bf 76}, 104032 (2007))
and   provides a new matrix representation for  cosets forming the
corresponding sigma-models in both cases.  Here we further improve
these methods introducing a matrix-valued dualisation procedure
which helps to avoid difficulties associated with solving the
dualisation equations in the component form. This new approach is
used to generate a five-parametric rotating charged Kaluza-Klein
black hole  with the squashed horizon adding one parameter more to
the recent solution by Tomizawa, Yasui and Morisawa which was
constructed using the previous version of the $G_{2(2)}$ generating
technique.
\end{abstract}


\maketitle

\section{Introduction}
Although all supersymmetric solutions solutions to minimal
five-dimensional supergravity were classified \cite{gaun}, to find
non-BPS configurations is a more difficult problem which attracts
attention in view of the discovery of black rings and their
generalizations \cite{empare}. An efficient tool to construct exact
solutions to multidimensional Einstein equations consists in
dimensional reduction based on the assumption of existence of a
sufficient number of commuting Killing symmetries (toroidal
reduction). Staring with D-dimensional Einstein equations coupled to
scalar and vector fields and assuming the existence of $D-3$ Killing
vectors one is able to derive three-dimensional gravity coupled
sigma-models in which the target scalar fields incorporate the
initial scalars, vectors and moduli of the toroidal reduction. With
some luck, the sigma-model target space happens to be a coset space
$G/H$ where $G$ is a semi-simple group known as the hidden symmetry
group (for a recent review see \cite{gerard}). The hidden symmetry
can be used directly to generate new solution form known ones with
the same three-dimensional metric. Another way of using the
dimensional reduction consists in solving the sigma-model equations
on the some submanifold of the target space (e.g. the null geodesic
method \cite{gerard} which is an alternative way to find BPS
solutions). Finally, one can  use the integrability methods assuming
one more Killing symmetry which allows  to construct two-dimensional
systems (for a concise formulation see, e.g., \cite{ga}).

Sigma-model technique for minimal five-dimensional supergravity was
developed in \cite{bccgsw},\cite{clem}, for an earlier discussion of
hidden symmetries in this theory see
\cite{mizo,cjlp,mizo2,pos1,pos,gnpp}. The hidden symmetry is this
case is the non-compact version $G_{2(2)}$ of the lowest exceptional
group $G_2$. To formulate the solution generating technique one has
to use some matrix representation of the coset. Representing the
seed solution in the matrix terms and acting by hidden symmetry
transformations  one can extract the sigma-model variables for  new
solutions. In \cite{bccgsw},\cite{clem} an   explicit $7\times 7$
representation of the $G_{2(2)}$ was used found earlier by Gunyadin
and Gursey \cite{gugu}.

The generalization to the case  of five-dimensional supergravity
with three $U(1)$ vector and two scalar fields was given in
\cite{GS}.   Apart from being more general, this theory is
interesting by the fact that the corresponding hidden symmetry is
given by such a familiar group as $SO(4,4)$ whose structure is
simpler than that of $G_{2(2)}$. Actually, one of the ways to
construct matrix representation of $G_{2(2)}$ consists in using
$SO(4,4)$ as a starting point \cite{gugu} and imposing some
constraints. It turns out that dealing with  the unconstrained
$SO(4,4)$ is in some sense easier. The matrix representation of the
relevant three-dimensional coset $SO(4,4)/SO(4)\times SO(4)$ was
given  in terms of the $8\times 8$ matrices split into the $4\times
4$ blocks. Freezing the scalar moduli  and identifying the vector
fields  reduces this theory back to minimal 5D supergravity thus
providing an alternative  to the technique of \cite{bccgsw}  in
terms of $8\times 8$ matrices.

Although construction of the new target space variables via hidden
symmetry transformations is a relatively simple problem, certain
complications may arise in solving the dualisation equations for the
one-form fields to scalars \cite{bccgsw,GS}. These  have to be
solved twice: for the seed solution to obtain its description in
terms of the coset matrix, and for the transformed solution in order
to extract the metric and the matter fields from the transformed
coset matrix. Though these equations are just linear partial
differential equations, solving them may present difficulties in the
case of complicated transformations. To remedy this problem, here we
propose to pass to dualized variables in the matrix form. Such a
possibility is suggested by the fact that the three dimensional dual
to the sigma-model matrix-valued current one-form is closed by
virtue of the equations of motion. Then locally it is an exact form,
and this provides the matrix-valued one-form whose exterior
derivative is dual to the the initial matrix current. This new
matrix transforms under the global action of the hidden symmetry by
some related  transformation, this allows us to find it
independently. From this matrix one can  read out the metric and
matter fields of the transformed solution algebraically, thus
avoiding the inverse dualisation problem.

We apply this approach to generate new Kaluza-Klein  squashed black
holes. Such black holes look as five-dimensional near the event
horizon possessing the $S^3$ angular section, but asymptotically
$S^3$ collapses to a twisted bundle of  $S^1$ over the base space
$S^2 $ with a constant radius of $S^1$ and growing radius of $S^2$.
Thus they become four-dimensional objects at infinity with a
compactified fifth dimension. Such a solution to five-dimensional
Einstein-Maxwell system was proposed by Ishihara and Matsuno
\cite{IM} in the non-rotating case. Its physical parameters and
thermodynamical properties were investigated in \cite{Ca}. Black
objects with such a topology is another alternative to usual
five-dimensional black holes and black rings. A certain class (but
not all) of squashed black holes can be obtained be the so-called
squashing transformation. This procedure was applied to
asymptotically flat \cite{IM,wang,NIMT} and non-asymptotically flat
solutions such as Kerr-G\"{o}del black holes \cite{TIMN,MINT,TI}. In
an attempt to enlarge the class of solutions, more recently
Tomizawa, Yasui and Morisawa \cite{tym} applied $G_{2(2)}$
transformations of \cite{bccgsw} to construct a generalization of
the charged Rasheed black hole \cite{rash} obtaining  a new solution
with four independent parameters: mass, angular momentum,
Kaluza-Klein parameter $\beta$ (in the notation of \cite{rash}) and
an electric charge. In this paper we will derive a more general
five-parametric solution adding as an independent parameter the
quantity $\alpha$ of \cite{rash}, which corresponds to an electric
charge in the four-dimensional interpretation of the Rasheed
solution.

The paper is organized as follows. In the following section we give
the brief review of the  derivation of the $SO(4,4)$ sigma-model.
Then (Sec. III) we introduce  the dualisation equations in the
matrix form and present the transformation law for the new
matrix-valued one-form.  In Sec. IV we explore which isometries
preserve the 5D Kaluza-Klein asymptotic behavior and describe the
strategy of the solution generation. Finally in Sec. V   we
demonstrate how the Rasheed solution can be obtained from the Kerr
metric using   $SO(4,4)$ transformations and construct a new
five-parametric squashed black hole.

\section{3D Sigma-model}
\subsection{Dimensional reduction}
In this section we briefly review  the derivation of the 3D
sigma-model for the $U(1)^3$ theory \cite{GS}. This theory may be
regarded as a truncated toroidal compactification of the 11D
supergravity:
 \be \label{ans11} I_{11} = \frac{1}{16\pi
G_{11}}\int\left(R_{11}\star_{11} \mathrm{1}-\frac12 F_{[4]}\wedge
\star_{11} F_{[4]} - \frac16F_{[4]}\wedge F_{[4]} \wedge
A_{[3]}\right),
 \ee
where $ F_{[4]} = dA_{[3]}$. Assuming an ansatz for the metric \be
ds_{11}^2 = ds^2_{5} + X^1 \left( dz_1^2 + dz_2^2 \right) + X^2
\left( dz_3^2 + dz_4^2 \right) + X^3 \left( dz_5^2 + dz_6^2
\right),\label{ds_11}\ee  and the form field  $$ A_{[3]} = A^1
\wedge dz_1 \wedge dz_2 + A^2 \wedge dz_3 \wedge dz_4 + A^3 \wedge
dz_5 \wedge dz_6,
$$ where all functions are independent of $z$, we obtain the
 the bosonic sector of 5D supergravity  coupled to three scalar
moduli $X^I$ ($I=1,2,3$), satisfying the constraint $X^1X^2X^3=1$,
and to three vector fields $A^I$:
 \bea\label{L5}
I_5 \!&=&\! \frac{1}{16 \pi G_5} \int\!\! \left( R_5 \star_5 1 \!-\!
\frac12 G_{IJ} dX^I \!\wedge\! \star_5 dX^J \!-\! \frac12G_{IJ} F^I
\!\wedge\! \star_5 F^J \!-\!
\frac{1}{6} \delta_{IJK} F^I \!\wedge\! F^J \!\wedge\! A^K \right)\!\!, \\
G_{IJ}&=&{\rm diag}\left((X^1)^{-2},\ (X^2)^{-2},\
(X^3)^{-2}\right),\quad F^{I}=dA^{I},\quad I,J,K=1,2,3.\nn
 \eea
Here the Chern-Simons coefficients $\delta_{IJK}=1$ for the indices
$ I,J,K $ being a permutation of 1, 2, 3, and zero otherwise.
Contraction of the above theory to minimal $5D$ supergravity is
effected via an identification of the vector fields:
$$A^1=A^2=A^3=\frac{1}{\sqrt{3}}A,$$ and freezing out  the moduli:
$X^1=X^2=X^3=1$. This leads to the Lagrangian  \be \mathcal{L}_5 =
R_5 \star_5 \mathbf{1} - \frac12 F\wedge \star_5 F -
\frac{1}{3\sqrt{3}} F\wedge F \wedge A.\nn \ee It is worth noting
that the 5D Einstein-Maxwell theory without the Chern-Simons term
does not lead to the three-dimensional sigma model with  a
semi-simple hidden symmetry group, so in this case the solution
generating technique can be formulated only for the static
truncation of the theory. This explains why the charged rotating
black hole solutions  are not known analytically.

Consider now further reduction to three dimensions assuming the
existence of two more commuting Killing symmetries. An overall
assumption for the 11D manifold will be ${\cal
M}_{11}=T^6\times\Sigma\times{\cal M}_3 $ where $\Sigma$ is $T^2$ if
both  these Killing vectors are asymptotically space-like, or
$T^1\times \mathbb{R}$ if one of them is asymptotically time-like.
The full set of 11D coordinates $x^{\mu},\ \mu=1,\ldots,11$ is thus
split into $z^a\in T^6,\ a=1,\ldots,6$, $x^i\in {\cal M}_3,\
i=1,\ldots,3$ and $z^p\in \Sigma,\ p=7,8$. The decomposition of the
5D metric  is given by
 \be
  ds_5^2=\lambda_{pq}(dz^p+a^p)(dz^q+a^q)-\kappa\tau^{-1}h_{ij}dx^i
 dx^j,\label{ds_5}
 \ee
where all metric functions are independent on $z^a$ and $z^p$.  The
5D metric components are parameterized by the KK one-forms
$a^p=a^p_idx^i$, the three-dimensional metric $h_{ij}$ of  ${\cal
M}_3$, three moduli fields $X^I,\ I=1,2,3$ and the scalars
$\varphi_1,\varphi_2,\chi$, which are arranged  in the following
$2\times 2$ matrix
 $$
 \lambda=\e^{-\frac{2}{\sqrt3}\varphi_1}\left( \begin{array}{cc}
1&
\chi \\
\chi & \chi^2+\kappa \e^{\sqrt{3}\varphi_1-\varphi_2}\end{array}
\right),\qquad  \det\lambda\equiv -\tau=\kappa
 \e^{-\frac{1}{\sqrt3}\varphi_1-\varphi_2}, \label{def_lambda}
 $$
where  $\kappa=\pm$ is responsible for the signature: $\kappa=1$ for
space-like $z^8$ , and $\kappa=-1$ for time-like $z^8$. The ansatz
(\ref{ds_11}) leads to the five-dimensional action (\ref{L5}). The
 5D $U(1)$ gauge fields $A^I$ reduce to the 3D one-forms
 $A^I(x^i)$ and the six axions collectively denoted as the
2D-covariant doublet
 $\psi_p^I=(u^I,v^I)$ with the index $p$ relative to the metric
 $\lambda_{pq}$
 $$
  A^I(x^i,z^7,z^8)=A^I(x^i)+\psi_p^I dz^p=A^I(x^i)+u^I dz^7+v^I
  dz^8.
 $$
 \subsection{Dualisation equations}
To obtain the three-dimensional sigma-model one has to dualize the
electro-magnetic (EM) one-forms $A^I$ and the KK one-forms $a^p$ to
scalars, which will be denoted as  $\mu_I$ and $\omega_p$. The
 dualisation equations read:
 \bea
  && \tau\lambda_{pq}da^q=\star V_p,\nn\\
  dA^I&=&d\psi_q^I\wedge a^q+\tau^{-1}G^{IJ}\star
  G_J,\label{eqs_of_dual}
 \eea
where the one-forms $G_I$ and $V_p$ are given by
  \bea
  &&G_I=d\mu_I+\frac12\delta_{IJK}d\psi_p^J
 \psi_q^K\varepsilon^{pq},\nn\\
  && V_p=d \omega_p- \psi_p^I  \Big( d \mu_I+\frac16 \delta_{IJK}d\psi_q^J
 \psi_r^K\varepsilon^{qr}\Big).\nn
  \eea
In the component form the Eqs.(\ref{eqs_of_dual}) read:\footnote{the
antisymmetrization is assumed with 1/2.}
 \bea
 &&\lambda_{pq}\partial^{{[}i} a^{j{]q}}=\frac{1}{2\tau \sqrt
 h}\varepsilon^{ijk}\Bigg[\partial_k \omega_p- \psi_p^I  \left(\partial_k \mu_I+
 \frac16 \delta_{IJK}\partial_k\psi_r^J
 \psi_t^K\varepsilon^{rt}\right)\Bigg],\nn \\
 &&\partial^{{[}i} A^{j{]}I}=a^{q{[}j}\partial^{i{]}}
 \psi_q^I+\frac{1}{2\tau\sqrt{h}}\varepsilon^{ijk}G^{IJ}\left(\partial_k\mu_J+
 \frac12\delta_{JKL}\partial_r\psi_p^K
 \psi_q^L\varepsilon^{pq}\right). \label{dualizequ}
 \eea
Substituting the   metric $ds_5^2$ in the form  (\ref{ds_5}) into
the 5D action (\ref{L5}) and performing   dualisation via
Eqs.(\ref{eqs_of_dual})  one derive the 3D gravity coupled
sigma-model:
 \be
I_3=\frac{1}{16\pi G_3}\int \sqrt{|h|}\left(R_3-{\cal
G}_{AB}\frac{\partial\Phi^A}{\partial
x^i}\frac{\partial\Phi^B}{\partial x^j}h^{ij}\right)d^3x,\label{L3}
 \ee
where the Ricci scalar $R_3$ is build using the 3-dimensional metric
$h_{ij}$. The set of potentials \footnote{The set
$\vec\phi=(\phi_1,\phi_2,\phi_3,\phi_4)$ comprises four scalars
related to  previously introduced  $\varphi_1,\varphi_2,\chi$ and
$X^I$ via
 \bea
  \phi_1&=&\frac{1}{\sqrt2}\left(-\ln(X^3)+\frac{1}{\sqrt3}\varphi_1+\varphi_2\right),\quad
   \phi_2=\frac{1}{\sqrt2}\left(\ln(X^3)-\frac{1}{\sqrt3}\varphi_1+\varphi_2\right),\nn
 \\
 \phi_3&=&\frac{1}{\sqrt2}\left(\ln(X^3)+\frac{2}{\sqrt3}\varphi_1\right),\qquad\qquad
 \phi_4=\frac{1}{\sqrt2}\,\ln\frac{X^1}{X^2}.\nn
 \eea} $\Phi^A=(\vec\phi,  \psi^I,
\mu_I,\chi,\omega_p),$ $A,B=1,\ldots,16$ realizes the harmonic map
$\Phi^A:\ x^i\in {\cal M}_3\ \rightarrow\ \Phi^A(x^i)\in {\cal
M}_{scal}$ between the 3D space-time ${\cal M}_3$ and the target
space (TS) ${\cal M}_{scal}$ with the metric ${\cal
G}_{AB}(\Phi^C)$. The target space line element
$dl^2={\cal{G}}_{AB}d\Phi^Ad\Phi^B$ has the form
 \bea
 dl^2& =& \frac12 G_{IJ}(dX^IdX^J+d{{\psi^I}^T}\lambda^{-1}
d\psi^J)-\frac12\tau^{-1}G^{IJ}G_IG_J+\frac14 \mathrm{Tr} \left(
\lambda^{-1} d\lambda \lambda^{-1} d\lambda \right)\nn\\
 & +& \frac14\tau^{-2} d\tau^2 - \frac12\tau^{-1} V^T \lambda^{-1}
V.\label{TS_metric}
 \eea
It is invariant under the action of the 28-parametric isometry group
$SO(4,4)$. The target space manifold ${\cal M}_{scal}$ is isomorphic
to the coset ${\cal M}=SO(4,4)/H$, where the isotropy group $H$ is
$SO(4)\times SO(4)$ for $\kappa=1$ and  $SO(2,2)\times SO(2,2)$ for
$\kappa=-1$. That is there is an isomorphic map $\pi$:
$\Phi^A\rightarrow \pi(\Phi^A)\in {\cal M}$. Moreover if $g\in
SO(4,4)$ is some constant element of the isometry group then the
following transformations
$$
 \pi \rightarrow \pi'=g\circ\pi,\quad ds^2_3\rightarrow ds^2_3
$$
leave invariant the action (\ref{L3}).
\subsection{Matrix representation}
As a convenient representative of the  coset $\pi(\Phi^A)\in {\cal
M}$ one can choose the matrix representation $\gamma: \pi
\rightarrow \gamma(\pi)\equiv {\cal V}$, where ${\cal V}$ is the
upper triangular matrix. We assume that ${\cal V}$ transforms under
the global action of the symmetry group $SO(4,4)$ by the right
multiplication and under the local action of the isotropy group $H$
by the left multiplication: ${\cal V}\to {\cal V}'=h(\Phi){\cal
V}g$, where $g$ and $h$ belong to the matrix representation $\gamma$
of $SO(4,4)$ and $H$ respectively. Given this representative, one
can construct the $H-$invariant matrix (which we denote the same
symbol ${\cal M}$ as the coset space)
 $$
 {\cal M}={\cal V}^T K {\cal V},
 $$
where $K$ is an involution matrix invariant under $H$: $h(\Phi)^T K
h(\Phi)=K,\label{h_K_h}$ and dependent on the coset signature
parameter $\kappa$. Then the transformation of the matrix ${\cal M}$
under $SO(4,4)$ will be
 \be
 {\cal M}\to {\cal M}'=g^T {\cal M}g.\label{trans_M}
 \ee
The target space metric (\ref{TS_metric}) in terms of the matrix
${\cal M}$ will read  \be
 dl^2=-\frac18 \mathrm{Tr}(d{\cal M}d{\cal M}^{-1}).\label{TS_by_M}
 \ee
Choosing suitable $8\times 8$ matrix representation $\gamma$ of the
isometry group $SO(4,4)$ we construct (see \cite{GS} for details )
the matrix representation of the coset ${\cal M}$ in terms of the
$4\times 4$ block matrices ${\cal P}={\cal P}^T$ and ${\cal
Q}=-{\cal Q}^{\widehat T}\footnote{$\widehat T$ denotes
transposition with respect to the minor diagonal}$ as follows
$$
 {\cal M}=\left( \begin {array}{cc}
 {\cal P}&{\cal P}{\cal Q}\\
 {\cal Q}^T{\cal P}&\widetilde{\cal P}+{\cal Q}^T{\cal P}{\cal Q}
 \end {array} \right),
 $$
 where the block matrices are given explicitly in the Appendix.
\section{Dualisation in the matrix form} As we have discussed, the
dualisation equations  (\ref{dualizequ}) may present difficulties in
applications of the solution generating technique. We can improve
the situation performing dualisation in the matrix form. Introducing
the matrix-valued current one-form  ${\cal J}$
$$
{\cal J}={\cal J}_idx^i={\cal M}d{\cal M}^{-1}
$$
we can rewrite the 3-dimensional sigma-model action (\ref{L3}) in
the following form
$$
I_3=\frac{1}{16\pi G_3}\int \left(R_3\star 1-\frac18
\mathrm{Tr}({\cal J}\wedge\star{\cal J})\right).
$$
In this expression the Hodge dual $\star\ $ is assumed with respect
to the 3-dimensional metric $h_{ij}$. Variation of this action with
respect to ${\cal J}$ shows that the two-form $\star{\cal J}$ is
closed: \be d\star{\cal J}=0.\label{eqs_of_motion}\ee Variation with
respect to the metric leads to  three-dimensional Einstein
equations: \be (R_3)_{ij}=\frac18 \mathrm{Tr}({\cal J}_i{\cal
 J}_j). \ee
 The first equation (\ref{eqs_of_motion}) means that the
 matrix-valued
 two-forms $\star {\cal
 J}$ is locally exact, i.e., it can be  presented
 as the exterior derivative of some matrix-valued one-form ${\cal
 N}$, that is
 \be
  \star{\cal J}={\cal M}\star d{\cal M}^{-1}=d{\cal N}.\label{def_N}
 \ee
 The matrix ${\cal N}$ is defined up to  adding an arbitrary matrix-valued
  closed one-form, which can be determined by choosing suitable
 asymptotic conditions. Now comparing the matrix dualisation equation
 (\ref{def_N}) with the initial dualisation equations
 (\ref{dualizequ})
we find  the following purely algebraic relations between certain
 components of the matrix $({\cal N})_{ab},\ a,b=1,\ldots, 8$ are
and the previous variables $a^p$ and
  $A^I$, namely
 \bea
  && a^7=({\cal N})_{16},\quad a^8=({\cal N})_{17},\nn\\
  && A^{1}=\psi^1_{p}a^p+({\cal N})_{15},\quad A^{2}=\psi^2_{p}a^p+({\cal
  N})_{14},\quad A^{3}=\psi^3_{p}a^p-({\cal
  N})_{26}.\label{rel_for_N}
 \eea
 Thus, if one manages to find the matrix ${\cal N}$, the metric and
 matter fields can be extracted algebraically.

 For the following it is important that the definition (\ref{def_N}) and
 the transformation law for the matrix ${\cal M}$ (\ref{trans_M})
 under the global transformations $g\in SO(4,4)$ imply the
 following global transformation of the matrix ${\cal N}$:
 $$
  {\cal N}\rightarrow {\cal N}\ '=g^T{\cal N}(g^T)^{-1}.
 $$

\section{Solution generating technique}
The sigma-model presented in the previous sections gives rise to
 generating technique which allows to construct new solutions
 from the known ones.  Let the metric $h_{ij}$ and the set of
potentials $\Phi^A$ combined in the coset matrix ${\cal M}$
correspond to the metric   and the three-form  of some 11D seed
solution. One has to extract part of the TS potentials from the seed
solution algebraically and solve the differential dualisation
equations (\ref{eqs_of_dual}) to find the remaining potentials.
Using the action of the TS isometries one can then construct a new
solution of the sigma-model with the same three-metric
$h_{ij}'=h_{ij}$ and the coset matrix
$$
 {\cal M}'=g^T{\cal M}g\quad (\hbox{or }{\cal M}'=g{\cal
 M}g^T),\quad g\in SO(4,4).
$$
Note that  five TS variables $\phi_1,\,\phi_2,\,\phi_3,\,\phi_4,\,
\chi$ enter the eleven-dimensional metric algebraically, via the
moduli $X^I,\, \lambda_{pq}$:
$$
  ds_{11}^2 = \sum_{I,a,a'} X^I \left( (dz^a)^2 + (dz^{a'})^2
  \right)+
  \lambda_{pq}(dz^p+a^p)(dz^q+a^q)+\tau^{-1}h_{ij}dx^idx^j,\ aa'=(12,34,56),
$$
while the KK vectors $a^p$ in the $T^2$ sector are related to the TS
potentials $\omega_p$ via dualisation. Similarly, in the form-field
sector,
$$
  A_{[3]}=(A^1+\psi_p^1 dz^p)\wedge dz^1 \wedge dz^2+
 (A^2+\psi_p^2 dz^p)\wedge dz^3 \wedge dz^4+
 (A^3+\psi_p^3 dz^p)\wedge dz^5 \wedge dz^6
$$
the six quantities $\psi_p^I$ are the TS potentials, while the
remaining one forms $A^I$ are related to the potentials $\mu_I$ via
dualisation. So the set of transformed potentials $\lambda_{pq}',\
(X^{I})'$ and $(\psi_p^{I})'$ can be explicitly extracted from the
coset matrix ${\cal M}'$. The remaining components of the
transformed metric $(ds_{11}^2)'$ and the 3-form $(A_{[3]})'$ which
are parametrized as the KK one-forms $(a^p)'$ and the EM fields
$(A^I)'$ are determined by the dualisation equations
(\ref{eqs_of_dual}). The inverse dualisation via the Eqs.
(\ref{dualizequ}) may be very difficult technically. Fortunately,
this problem can be reduced to a purely algebraic one using the
dualisation in the matrix form (\ref{def_N}) as described in the
previous section. Taking into account that  the matrix ${\cal N}$
transforms as
$$
 {\cal N}\ '=g^T{\cal N}(g^T)^{-1}\quad (\hbox{or } {\cal N}\ '=g{\cal N}g^{-1}\
 ),\quad g\in SO(4,4)
$$ and using the relations (\ref{rel_for_N}) one can easily obtain the
desired quantities $(a^p)'$ and $(A^I)'$.

We will denote the 28 generators of the $so(4,4)$ algebra as $${\cal
T}=(H_1,H_2,H_3,H_4,\ P^{\pm I},\, W_{\pm I},\, Z_{\pm
I},\,\Omega^{\pm p},\,X^{\pm}),$$ with $I= 1,2,3,\, p=7,8$. Their
matrix representation   can be found in the Appendix. The
corresponding one-parametric transformations $g=\e^{\alpha{\cal
T}}$, where $\alpha$ is a transformation parameter, give the set of
the target space isometries.
\subsection{Asymptotic conditions} An important question is how to
identify the isometries we need to use in order to construct
solutions with the desired properties. These are usually associated
with asymptotic conditions. In this paper we consider asymptotic
conditions corresponding to 5D Kaluza-Klein black holes with
squashed horizons  embedded into eleven dimensions which correspond
to the following asymptotic manifold: $T^6\times
\mathbb{R}^{1}\times S_{sq}$, where $S_{sq}$ is a squashed $S^3$. We
will assume that TS potentials have the following general asymptotic
behavior
 \be
  \lambda\sim\left(\begin{array}{cc}
  1 & 0 \\
  0 & -1 \\\end{array}\right) +\frac{\delta\lambda}{r},\quad
  \omega_7\sim \frac{\delta\omega_7}{r},\quad \omega_8\sim
 \frac{\delta\omega_8}{r^2}, \quad A_{[3]}=0,\label{asym_cond}
  \ee
 where $\delta\lambda,\ \delta\omega_7$ and $\delta\omega_8$ are
 constant.
The asymptotic behavior with
$\delta\lambda=\delta\omega_7=\delta\omega_8=0$ correspond to the
trivial $S^1$ bundle over a 4D Minkowski space-time. The asymptotic
coset matrix for this case is ${\cal M}_{as}=K$ which is preserved
under the isometries belonging to the isotropy group $H$ of the
$SO(4,4)$:
$$
P^I+ P^{-I},\quad Z_I+ Z_{-I},\quad W_I-W_{-I},\quad X^{+}+
 X^{-},\quad \Omega^7+ \Omega^{-7},\quad
 \Omega^8-\Omega^{-8}.
$$
For more general asymptotic behavior such as (\ref{asym_cond}) one
have use the above transformations with some constraints on the
parameters.

To apply these isometries in the case of minimal 5D supergravity one
needs to find the relevant embedding of the $G_{2(2)}$ subgroup into
 $SO(4,4)$. As was shown in \cite{GS}, the following combinations of the
$SO(4,4)$ generators realize the positive and negative root
generators of $G_{2(2)}$:
$$
 P^{\pm}\sim \sum P^{\pm I},\quad Z_{\pm}\sim \sum Z_{\pm I},\quad
 W_{\pm}\sim \sum W_{\pm I},\quad \Omega^{\pm p},\quad X^{\pm}.
$$
Thus the isometries
$$
P^{+}+ P^{-},\quad Z_{+}+ Z_{-},\quad W_{+}-W_{-},\quad X^{+}+
 X^{-},\quad \Omega^7+ \Omega^{-7},\quad
 \Omega^8-\Omega^{-8}
$$
can be used to generate  new KK solutions in the minimal 5D
supergravity.
\section{Constructing five-parametric squashed black hole}
\subsection{From Kerr  to Rasheed solution}
First we would like to demonstrate how to construct using our
technique the rotating dyonic black hole of \cite{rash} from the
Kerr metric. We define the coordinates  $z^7=x^5,\ z^8=t$ and
$x^i=(r,\theta,\phi)$. In this basis the Kerr solution of the mass
$M_K$ and the angular momentum $J_K=aM_K$ smeared into the fifth
dimension reads
$$
 ds^2_5=(dx^5)^2-(1-Z)\Bigl(dt+\frac{aZ\sin^2\theta }{1-Z}d\phi\Bigr)^2+\frac{\rho}{\Delta}dr^2
 +\rho d\theta^2+\frac{\Delta}{1-Z}\sin^2\theta d\phi^2,
$$
where
$$
 \rho=r^2+a^2\cos^2\theta,\quad \Delta=r^2-2M_Kr+a^2,\quad
 Z=\frac{2M_Kr}{\rho}.
$$
The corresponding TS variables are:
 \bea
 &&\lambda_{pq}=\left(\begin {array}{cc}
 1&0\\
 0&Z-1\end {array} \right),\quad \tau=1-Z,\nn\\
 &&\omega_7=0,\quad \omega_8=\frac{2M_K a\cos\theta}{\rho},\quad \Bigl(
 a^7_{\phi}=0,\quad a^8_{\phi}=\frac{aZ\sin^2\theta}{1-Z}\Bigr).\nn
 \eea
 The above definitions of the TS potentials lead to the following
  blocks of the coset matrix
 ${\cal M}$
 $$
  {\cal Q}=\left(\begin {array}{cccc}
 0&0&\frac{2M_K a\cos\theta}{\rho}&0\\
 0&0&0&0\\
 0&0&0&0\\
 0&0&0&0 \end {array}\right),\quad {\cal P}=
 \left(\begin {array}{cccc}
 \frac{1}{Z-1}&0&0&0\\
 0&\frac{1}{Z-1}&0&0\\
 0&0&1&0\\
 0&0&0&1 \end {array}\right).
 $$
One can easily obtain the dual matrix  ${\cal N}$ solving the
Eq.(\ref{def_N}) :
$$
 {\cal N}=\left(\begin {array}{cccccccc}
 -\frac{2M_K\Delta\cos\theta}{\rho(1-Z)}&0&0&0&0&0&\frac{Za\sin^2\theta}{1-Z}&0\\
 0&-\frac{2M_K\Delta\cos\theta}{\rho(1-Z)}&0&0&0&0&0&-\frac{Za\sin^2\theta}{1-Z}\\
 0&0&0&0&0&0&0&0\\
 0&0&0&0&0&0&0&0\\
 0&0&0&0&0&0&0&0\\
 0&0&0&0&0&0&0&0\\
 -\frac{Za(r-2M_k)\sin^2\theta}{r(1-Z)}&0&0&0&0&0&\frac{2M_K\Delta\cos\theta}{\rho(1-Z)}&0\\
 0&\frac{Za(r-2M_k)\sin^2\theta}{r(1-Z)}&0&0&0&0&0&\frac{2M_K\Delta\cos\theta}{\rho(1-Z)}\end {array}\right)d\phi.
$$
The dyon solution of \cite{rash}  was generated by applying the
constrained $SO(1,2)$ transformations to the smeared  Kerr solution.
The constraint ensures  asymptotic flatness and absence of the NUT
parameter in the four-dimensional solution. The latter requirement
means that the asymptotic of the $\phi$-component of the KK one-form
$a^8$  decays as $O(\frac{1}{r})$ after  applying the
transformation. This leads to one relation between three parameters
of the general $SO(1,2)$ transformation so only two parameters
remain independent.

The $\sigma$-model with the $SO(4,4)$ isometry group reduces to that
of the five-dimensional KK gravity (embedded into 11D) with the
$SL(3,\mathbb{R})$ isometry group under the conditions $X^I=1,\
A_{[3]}=0$. So there exists the $SO(1,2)$ subgroup of the $SO(4,4)$
group which produces a dyonic rotating KK black hole. One finds that
the this  subgroup is generated by the elements $X^{+}+ X^{-},\
\Omega^7+ \Omega^{-7},\ \Omega^8-\Omega^{-8}$.

We  perform  the transformation $g=g_1g_2g_3,\ g\in SO(1,2)$, where
$$
 g_1=\e^{\alpha(X^{+}+X^-)},\quad
 g_2=\e^{\beta(\Omega^7+\Omega^{-7})},\quad
 g_3=\e^{\gamma(\Omega^8-\Omega^{-8})},
$$
 assuming that the matrices ${\cal M}$ and ${\cal N}$ are transformed under
$g$ as ${\cal M}_1=g{\cal M}g^T$ and ${\cal N}_1=g{\cal N}g^{-1}$
respectively. Then we demand that $g$ preserve the $O(\frac{1}{r})$
asymptotic behavior of $a^8_{\phi}$ or, equivalently, the
$O(\frac{1}{r^2})$ asymptotic behavior of $\omega_8$. This give the
same relation between three parameters $\alpha,\beta,\gamma$   as in
\cite{rash}:
$$
 \tan2\gamma=\tanh\alpha\sinh\beta.
$$
The metric of \cite{rash} now can be extracted from the transformed
matrices ${\cal M}_1$ and ${\cal N}_1$ as usual.  But for further
application of other
 isometries of the TS we need to know only the matrices ${\cal M}_1$
and ${\cal N}_1$.

\subsection{Charging the Rasheed solution}
Our improved generating technique allows us  to construct the
charged Rasheed solution without the restriction $\alpha=0$
(corresponding to absence of the electric charge $Q$ in the
four-dimensional interpretation of the solution \cite{rash}) which
was imposed in \cite{tym} apparently because of the difficulties
with inverse dualisation. Using our Eq. (\ref{eqs_of_dual}) we   can
extract the finial solution from the target space potentials purely
algebraically.

To obtain the charged solution from the vacuum Rasheed solution  we
apply   the following global transformation with the parameter
$\delta$ to the matrices ${\cal M}_1$ and ${\cal N}_1$:
$$
 {\cal M}_2=\Pi{\cal M}_1\Pi^T,\quad {\cal N}_2=\Pi{\cal
 N}_1\Pi^{-1},\quad \Pi=\e^{\delta\sum_{I} (Z_I+Z_{-I})}.
$$
As was shown in \cite{GS}, the transformation $\Pi$ is equivalent to
the action of the one-parametric $G_{2(2)}$ subgroup, which generate
an electric charge. Then extracting the TS variables from ${\cal
M}_2$ one can find  the transformed potentials $\psi^I_p{}'$ and
$\lambda_{pq}'$:
 \bea
&&v'=v_1'=v_2'=v_3'=csD^{-1}\Bigl(({\cal M}_1)_{11}+({\cal
M}_1)_{33}\Bigr),\nn\\
&&u'=u_1'=u_2'=u_3'=-csD^{-1}\Bigl(s({\cal M}_1)_{17}+c({\cal
M}_1)_{23}\Bigr),\nn\\
&&\lambda_{88}'=D^{-2}({\cal M}_1)_{11}({\cal M}_1)_{33},\nn\\
&&\lambda_{78}'=-{\lambda_{88}'}\Biggl(c^3\frac{({\cal M}_1)_{23}}
{({\cal M}_1)_{33}} -s^3\frac{({\cal M}_1)_{17}}{({\cal
M}_1)_{11}}\Biggr),\nn\\
&&\lambda_{77}'=\frac{D}{({\cal M}_1)_{11}({\cal
M}_1)_{33}}+\frac{({\cal M}_1)_{11} ({\cal
M}_1)_{33}}{D^2}\Biggl(c^3\frac{({\cal M}_1)_{23}} {({\cal
M}_1)_{33}}
-s^3\frac{({\cal M}_1)_{17}}{({\cal M}_1)_{11}}\Biggr)^2,\nn\\
&&\tau'=-D^{-1},\quad (X^I)'=1.\nn
 \eea
where $c=\cosh\delta,\ s=\sinh\delta$ and
$$
 D=c^2({\cal M}_1)_{11}+s^2({\cal M}_1)_{33}.
$$
The explicit  expressions for the coset components $({\cal
M}_1)_{ab},\ a,b=1,\ldots,8$ are
 \bea
 ({\cal M}_1)_{11}&=&\frac{1}{p(Z-1)}\Biggl\{p+\frac{2M_K}{\rho}
 \Bigl(a\cos\theta
 s_{\alpha}c_{\beta}^2s_{\beta}+r(c_{\beta}^2c_{\alpha}-p)-M_Kc_{\beta}^2(c_{\alpha}-p)\Bigr)\Biggr\},\nn\\
 ({\cal M}_1)_{33}&=&-\frac{1}{p(Z-1)}\Biggl\{p+\frac{2M_K}{\rho}\Bigl
 (a\cos\theta s_{\alpha}s_{\beta}(1+c_{\alpha}^2c_{\beta}^2)\nn\\
 &+&(r-M_K)\Bigl[c_{\alpha}(c_{\beta}^2-p^2)+p(c_{\alpha}^2c_{\beta}^2-1)\Bigr]-
 rpc_{\alpha}^2c_{\beta}^2\Biggr\},\nn\\
 ({\cal M}_1)_{23}&=&\frac{2M_K}{\rho(Z-1)}\Biggl\{c_{\alpha}s_{\beta}pa\cos\theta
 -rps_{\alpha}+M_Ks_{\alpha}(p-c_{\alpha}c_{\beta}^2)\Biggr\},\nn\\
 ({\cal
 M}_1)_{17}&=&\frac{2M_Kc_{\beta}}{\rho(Z-1)}\Biggl\{ap\cos\theta
 +s_{\alpha}s_{\beta}M_K\Biggr\},\nn
 \eea
 where
 $$
 p=\sqrt{c_\alpha^2+s_{\alpha}^2s_{\beta}^2},\quad
c_{\star}=\cosh\star,\ s_{\star}=\sinh\star.
  $$
Similarly, the relations (\ref{rel_for_N}) give the transformed KK
one-forms $(a^p)'$ and the five-dimensional one-form $A'$:
 \bea
&&(a^7)'=({\cal N}_1)_{16},\quad (a^8)'=c^3({\cal
N}_1)_{17}+s^3({\cal N}_1)_{32},\nn\\
&&A'=u'dx^5+v'dt+d\phi[u'(a_{\phi}^7)'+v'(a_{\phi}^8)'-cs\Bigl(c({\cal
N}_1^{\phi})_{17}+s({\cal N}_1^{\phi})_{32}\Bigr)]\nn,
  \eea
where
 \bea
({\cal N}_1)_{16}&=&\frac{2M_K}{\rho (Z-1)p
}\Biggl\{ac_{\beta}s_{\alpha}\sin^2\theta\Bigl[p(r-M_K)+M_K
c_{\alpha}c_{\beta}^2\Bigr]
-\Delta s_{\beta}c_{\beta}\cos\theta\Biggr\}d\phi,\nn\\
({\cal N}_1)_{17}&=&-\frac{2M_K}{\rho (Z-1)
}\Biggl\{ac_{\beta}\sin^2\theta\Bigl[M_K p+c_{\alpha}(r-M_K)\Bigr]\Biggr\}d\phi,\nn\\
({\cal N}_1)_{32}&=&-\frac{2M_K}{\rho (Z-1)
}\Biggl\{as_{\beta}\sin^2\theta\Bigl[c_{\alpha}M_K
p+M_K-r\Bigr]+\Delta s_{\alpha}p\cos\theta\Biggr\}d\phi.\nn
 \eea
To write the resulting metric we introduce the  functions
$A,B,C,E,W,X,Y$ via
 \bea
 &&A=({\cal M}_1)_{11},\quad B=({\cal M}_1)_{33},\quad  C=({\cal
 M}_1)_{17},\quad E=({\cal M}_1)_{23},\nn\\
 &&W=({\cal N}_1)_{16}^{\phi},\quad X=({\cal N}_1)_{17}^{\phi},\quad Y=({\cal
 N}_1)_{32}^{\phi}. \nn
 \eea
In   terms of them the eleven-dimensional metric will read
 \bea
ds_{11}^2 &=& \sum_{a,a'} \left( (dz^a)^2 + (dz^{a'})^2
  \right)\nn\\
  &+&f(dt+\Omega')^2+\frac{1}{fD}(dx^5+Wd\phi)^2-D\left(\frac{\rho}{\Delta}dr^2
 +\rho d\theta^2+\frac{\Delta}{1-Z}\sin^2\theta d\phi^2\right), \nn
  \eea
with
 \bea
  &&f=\frac{AB}{D^2},\quad \Omega'=\Omega_5 dx^5+\Omega_\phi d\phi,\quad D=Ac^2+Bs^2\nn\\
  &&\Omega_5=\frac{C}{A}s^3-\frac{E}{B}c^3,\quad
  \Omega_\phi=\frac{WC+YA}{A}s^3+\frac{XB-WE}{B}c^3.\nn
  \eea
The corresponding 3-form $A_{[3]}$ can be written as
$$
 A_{[3]}=\sum_{a,a'}A'\wedge dz^a\wedge dz^{a'},
$$
where the five-dimensional one-form $A'$ reads
$$
 A'=\frac{cs}{D}\Bigl\{(A+B)dt-(sC+cE)dx^5+\Bigl[c(XB-WE)-s(WC+YA)\Bigr]d\phi\Bigr\}.
$$
Our new solution contains five free parameters $M_K, a , \alpha,
\beta, \delta$ and reduces to that of \cite{tym} if $\alpha=0$.
Physical properties of the new solution will be discussed in a
separate publication.
\section{Conclusions}
In this paper we have suggested an  improved solution generating
technique for the $5D\ U(1)^3$ supergravity based on the 3D
sigma-model with the $SO(4,4)$ isometry group. The main new
ingredient is the matrix dualisation equation which opens a way to
avoid the inverse dualisation problem in constructing new solutions
form old. As an application we have obtained the five-parametric
Kaluza-Klein black hole of the minimal 5D supergravity generalizing
the solution by Tomizawa, Yasui and Morisawa \cite{tym}.

Note that we did not use the most general transformations preserving
the desired asymptotic structure $T^6\times \mathbb{R}^{1}\times
S_{sq}$. Namely, the combination of the transformations
$\e^{\delta_1\sum_I (P^I+P^{-I})},\ \e^{\delta_2\sum_I
(Z_I+Z_{-I})}$ and $\e^{\delta_3\sum_I (W_I-W_{-I})}$ will give more
general solutions with the same asymptotic behavior provided some
constraint on the parameters $\delta_1,\delta_2,\delta_3$ holds.

Also, we have restricted attention here by the case of minimal 5D
supergravity. The corresponding generalization to the $U(1)^3$
theory is straightforward.

 \begin{acknowledgments}
The authors are grateful to Gerard Cl\'ement and Chiang-Mei-Chen for
helpful discussions. The paper was supported by the RFBR grant
08-02-01398-a.
 \end{acknowledgments}

\appendix

\section{$8\times 8$ matrix representation }
We choose the following $8\times 8$ matrix representation of the
so(4,4) algebra
 \be
  E= \left( \begin{array}{cc}
A & B \\
C & -A^{\widehat{T}} \end{array} \right),\label{def_matr_basis}
 \ee
 where
$A,\ B,\ C$ are the $4\times 4$ matrices, $A,\ B$ being
antisymmetric, $B=-B^{T},\ C=-C^{T}$, and the symbol $\widehat{T}$
in $A^{\widehat{T}}$ means  transposition with respect to the minor
diagonal. The diagonal matrices $\vec H$ are given by the following
$A-$type matrices (with $B=0=C$): \bea
 A_{H_1}\!\!&=&\!\!\left( \begin {array}{cccc}
 \sqrt{2}&0&0&0\\0&0&0&0\\0&0&0&0
 \\0&0&0&0\end {array} \right), \
 A_{H_2}\!=\!\left(\begin {array}{cccc} 0&0&0&0\\0&\sqrt{2}&0&0\\0&0&0&0
 \\0&0&0&0\end {array} \right),\
 A_{H_3}\!=\!\left(\begin {array}{cccc} 0&0&0&0\\0&0&0&0\\0&0&\sqrt{2}&0
 \\0&0&0&0\end {array} \right),\
 A_{H_4}\!=\!\left(\begin {array}{cccc} 0&0&0&0\\0&0&0&0\\0&0&0&0
 \\0&0&0&\sqrt{2}\end {array} \right).\nn
 \eea
Twelve generators corresponding to the positive roots are given by
the upper-triangular matrices $E_k,\ k=1,\ldots,12, $.  From these
the generators labeled by $k=2,4,6,7,9,12$ are of pure $A$-type
(with $B=0=C$):
 \bea
 &&A_{E_2}=\left(\begin {array}{cccc} 0&0&0&1\\0&0&0&0\\0&0&0&0
 \\0&0&0&0\end {array} \right),\quad
 A_{E_4}=\left(\begin {array}{cccc} 0&0&0&0\\0&0&0&0\\0&0&0&-1
 \\0&0&0&0\end {array} \right), \quad
 A_{E_6}=\left(\begin {array}{cccc} 0&1&0&0\\0&0&0&0\\0&0&0&0
 \\0&0&0&0\end {array} \right),\nn\\
 &&A_{E_7}=\left(\begin {array}{cccc} 0&0&0&0\\0&0&0&-1\\0&0&0&0
 \\0&0&0&0\end {array} \right)\quad
 A_{E_9}=\left(\begin {array}{cccc} 0&0&-1&0\\0&0&0&0\\0&0&0&0
 \\0&0&0&0\end {array} \right),\quad
A_{E_{12}}=\left(\begin {array}{cccc} 0&0&0&0\\0&0&-1&0\\0&0&0&0
 \\0&0&0&0\end {array} \right).\nn\eea
 while the other six are of pure $B$ type (with $A=0=C$):
 \bea &&B_{E_1}=\left(\begin {array}{cccc} 1&0&0&0\\0&0&0&0\\0&0&0&0
 \\0&0&0&-1\end {array} \right),\quad
 B_{E_3}=\left(\begin {array}{cccc} 0&0&0&0\\0&-1&0&0\\0&0&1&0
 \\0&0&0&0\end {array} \right),\quad
 B_{E_5}=\left(\begin {array}{cccc} 0&0&0&0\\0&0&0&0\\-1&0&0&0
 \\0&1&0&0\end {array} \right),\nn\\
 &&B_{E_8}=\left(\begin {array}{cccc} 0&0&0&0\\-1&0&0&0\\0&0&0&0
 \\0&0&1&0\end {array} \right),\quad
 B_{E_{10}}=\left(\begin {array}{cccc} 0&1&0&0\\0&0&0&0\\0&0&0&-1
 \\0&0&0&0\end {array} \right),\quad
 B_{E_{11}}=\left(\begin {array}{cccc} 0&0&1&0\\0&0&0&-1\\0&0&0&0
 \\0&0&0&0\end {array} \right).\nn
 \eea
The correspondence with the previously introduced generators is as
follows ($I=1,2,3,\;p=7,8$): \be P^I\leftrightarrow {E_I}, \quad
W_I\leftrightarrow { E_{I+3}}, \quad Z_I\leftrightarrow {E_{I+6}},
\quad \Omega^p\leftrightarrow {E_{p+3}},\quad X^{+} \leftrightarrow
{E_{12}}.\nn\ee In this representation, the matrices corresponding
to the negative roots, \be P^{-I}\leftrightarrow E_{-I}, \quad
W_{-I}\leftrightarrow E_{-(I+3)}, \quad Z_{-I}\leftrightarrow
E_{-(I+6)}, \quad \Omega^{-p}\leftrightarrow E_{-(p+3)},\quad X^{-}
\leftrightarrow E_{-12},\nn\ee are  transposed with respect to the
positive roots matrices:  \be
 E_{-k}=(E_k)^T.\nn
 \ee
The following normalization conditions are assumed: \be
 \mathrm{Tr}(H_i,H_j)=4\delta_{ij},\ i,j=1\ldots 4,\qquad
 \mathrm{Tr}(E_k,E_{-k})=2,\nn
 \ee
and the involution matrix $K$   is chosen as \be
 K=\rm{diag}(\kappa,\kappa,1,1,1,1,\kappa,\kappa).\nn
 \ee
The generators of the isotropy subgroup are selected by the equation
$h(\Phi)^T K h(\Phi)=K$. They are given by the following linear
combinations of the generators: \be
 P^I-\kappa P^{-I},\quad Z_I -\kappa Z_{-I},\quad W_I-W_{-I},\quad
 X^{+}-\kappa X^{-},\quad \Omega^7-\kappa \Omega^{-7},\quad
 \Omega^8-\Omega^{-8}.
\nn \ee
\section{Matrix representation of coset ${\cal M}$}
$$
 {\cal M}=\left( \begin {array}{cc}
 {\cal P}&{\cal P}{\cal Q}\\
 {\cal Q}^T{\cal P}&\widetilde{\cal P}+{\cal Q}^T{\cal P}{\cal Q}
 \end {array} \right),
 $$
 where the $4\times 4$ blocks ${\cal P}$ and ${\cal Q}$ are
\bea
 &&{\cal Q}=\left( \begin{array}{cccc}
 \mu_1+ \frac{u^3v^2-v^3u^2}{2}&\omega_7-
 \frac{u^3v^1u^2-2u^3v^2u^1+v^3u^1u^2}{6}-u^2\mu_2,&
 \omega_8+\frac{v^3u^1v^2-2v^3u^2v^1+u^3v^1v^2}{6}-v^2\mu_2&0\\
  -v^2&-\mu_3+\frac{v^1u^2-u^1v^2}{2}&0&\\
   -u^2&0&&\\
    0&&&\\
 \end{array} \right),\nn\\
 &&{\cal P}=\left( \begin {array}{cc}
 \Psi^T\Lambda \Psi,&\Psi^T\Lambda\Phi\\
 \Phi^T\Lambda\Psi,&\Phi^T\Lambda\Phi+\e^{\sqrt2\phi_4}
 \end {array} \right),\label{def_P}\quad \widetilde{\cal P}=({\cal
 P}^{-1})^{\widehat T}.\nn
 \eea
 Here $\Psi$ and $\Lambda$ are the $3\times 3$ matrices
 $$
 \Psi=\left( \begin {array}{ccc}
 1&u^3&-v^3\\
 0&1&0 \\
 0&0&1 \end {array}
 \right),\quad
 \Lambda=\kappa\left( \begin {array}{ccc}
 \e^{\sqrt2\phi_1}&0&0\\
 0&\e^{\sqrt2\phi_2}&-\chi\e^{\sqrt2\phi_2} \\
 0&-\chi\e^{\sqrt2\phi_2}&\e^{\sqrt2\phi_2}\chi^2+\kappa\e^{\sqrt2\phi_3} \end {array}
 \right)
  $$
  and $\Phi$ is the 3-column
  $$
   \Phi=\left( \begin {array}{ccc}
 \mu_2+\frac12(u^1v^3-u^3v^1)\\
 -v^1\\
 -u^1\end {array}
 \right).
  $$


\end{document}